\documentclass{nature}
\usepackage[utf8]{inputenc}
\usepackage{graphicx}
\usepackage{float}
\usepackage{gensymb}
\usepackage{upgreek}
\usepackage{color,amsmath}
\usepackage[colorlinks,linkcolor=black,citecolor=black,urlcolor=blue]{hyperref}
\usepackage[normalem]{ulem}
\usepackage{siunitx}
\usepackage{enumitem}
\usepackage{ulem}
\usepackage{amsmath,amsfonts}
\usepackage{siunitx}
\usepackage{ragged2e}
\usepackage{textcomp}
\usepackage{color,soul}
\usepackage{hyphenat}
\usepackage[labelfont=bf]{caption}
\title{Magnetoelectric phase control at domain-wall-like \\ epitaxial oxide multilayers}

\author{Elzbieta Gradauskaite\textsuperscript{1,*}, Chia-Jung Yang\textsuperscript{1}, Shovon Pal\textsuperscript{1,2}, Manfred Fiebig\textsuperscript{1}, Morgan Trassin\textsuperscript{1,*} \\ \textsuperscript{*}elzbieta.gradauskaite@cnrs-thales.fr, morgan.trassin@mat.ethz.ch}

\makeatletter
\let\saved@includegraphics\includegraphics
\AtBeginDocument{\let\includegraphics\saved@includegraphics}
\renewenvironment*{figure}{\@float{figure}}{\end@float}

\makeatother

\DeclareUnicodeCharacter{2212}{-}

\begin{document}

\maketitle
\begin{affiliations}
 \item Department of Materials, ETH Zurich, Vladimir-Prelog-Weg 4, 8093 Zurich, Switzerland
 \item School of Physical Sciences, National Institute of Science Education and Research, An OCC of HBNI, Jatni, 752 050 Odisha, India
\end{affiliations}
\captionsetup{font={stretch=0.8}} 

\textbf{Ferroelectric domain walls are nanoscale objects that can be created, positioned, and erased on demand. They often embody functional properties that are distinct from the surrounding bulk material. Enhanced conductivity, for instance, is observed at charged ferroelectric domain walls. Regrettably, domain walls of this type are scarce because of the energetically unfavorable electrostatics. This hinders the current technological development of domain-wall nanoelectronics. Here we overcome this constraint by creating robust domain-wall-like objects in epitaxial oxide heterostructures. We design charged head-to-head (HH) and tail-to-tail (TT) junctions with two ferroelectric layers (BaTiO\textsubscript{3} and BiFeO\textsubscript{3}) that have opposing out-of-plane polarization. To test domain-wall-like functionalities, we insert an ultrathin ferromagnetic La\textsubscript{0.7}Sr\textsubscript{0.3}MnO\textsubscript{3} layer into the junctions. The interfacial electron or hole accumulation at the interfaces, set by the HH and TT polarization configurations, respectively, controls the LSMO conductivity and magnetization. We thus propose that trilayers reminiscent of artificial domain walls provide magnetoelectric functionality and may constitute an important building block in the design of oxide-based electronic devices.}

\justifying

\section{INTRODUCTION}

Oxide heterostructures exhibit a plethora of technologically relevant phenomena that arise at their interfaces\cite{Zubko2011a,Hwang2012}, and they offer structural versatility, resulting from the combination of different materials. Their atomic construction and related functionalities, however, are generally inflexible once their growth has been completed. Ferroelectric domain walls as a special type of quasi-two-dimensional objects are superior in this respect. Even post-growth they can be nucleated, moved, and erased on demand\cite{Catalan2012,Whyte2014,Sharma2017}. In particular, charged domain walls may exhibit conduction distinctly different from the bulk\cite{Meier2012}. This functionality is determined by the local polarization configuration set by the surrounding domains\cite{Meier2015}, which can be reversibly tuned by electric-field poling, and therefore is of particular interest for device applications. However, the occurrence of charged domain walls is impeded by the unfavorable electrostatics of a head-to-head (HH) or tail-to-tail (TT) meeting of electric-dipole moments at the walls\cite{Seidel2009,Meier2012}. Moreover, the properties of domain walls are limited by those of the host material. Here, merging the functionalities of oxide interfaces and ferroic domain walls by the construction of domain-wall-like epitaxial oxide multilayers would lead to the ``best of both worlds'', enabling robust and localized, yet widely tunable interfacial properties. 

The domain-wall-like tunability of conductivity can be brought to the oxide interfaces by considering a trilayer architecture in which a charge-sensitive metallic layer with competing electronic ground states is sandwiched between two ferroelectric layers. If the metal is (anti-)ferromagnetic, its spin ordering widens the perspective towards magnetoelectronics and spintronics. In this way, the trilayer emulates a functional magnetoelectric domain wall, as both the metal's conductivity and magnetization can be modulated by the orientation of the switchable spontaneous electric polarization in the adjacent layers through the associated effective charge doping. 

Such charge doping has been widely investigated in composite multiferroic heterostructures\cite{FernandesVaz2013a,Trassin2016a,Gradauskaite2021} by combining ferroelectric and ferromagnetic layers. The vast majority of experiments up to this day have, however, focused on multiferroic bilayers\cite{Molegraaf2009a,Jiang2012a, Yi2013, Ma2014a, Popescu2015, Meyer2016, Li2017} with only a few exceptions of superlattice studies\cite{Singh2014b, Guo2017b, Chen2018b}. In contrast, a trilayer design (ferroelectric\(|\)ferromagnet\(|\)ferroelectric) enables the stabilization of opposite polarization directions in the ferroelectric films. This is key to domain-wall-like HH and TT polarization configurations that offer coordinated charge doping at both interfaces of the ferromagnetic element. Unfortunately, the choice and stabilization of the polarization direction in each layer is difficult to achieve, as polarization in heterostructures can reorient spontaneously or decay into a multi-domain configuration. Furthermore, once capped with a metal, polar states can no longer be probed with established techniques like piezoresponse force microscopy (PFM). This has so far obstructed the realization of heterostructures emulating domain-wall functionalities.

Here we design multiferroic heterostructures with functional trilayers resembling charged ferroelectric domain walls with added magnetic functionality. We track the HH and TT configurations of polarization during the growth of our model BTO\(|\)LSMO\(|\)BFO and BFO\(|\)LSMO\(|\)BTO systems (where BTO is BaTiO\textsubscript{3}, LSMO is La\textsubscript{0.7}Sr\textsubscript{0.3}MnO\textsubscript{3}, BFO is BiFeO\textsubscript{3}), using in-situ second harmonic generation (ISHG). The corresponding hole and electron accumulations at the ferromagnetic LSMO layer trigger an interfacial magnetoelectric coupling. Using a combination of magnetometry with a non-invasive superconducting quantum interference device (SQUID) and terahertz time-domain spectroscopy (THz-TDS), we show that the artificial domain wall can exhibit different magnetic signatures and conduction properties, depending on its HH or TT polarization configuration. Notably, the LSMO layer in the HH configuration transitions from an insulating and paramagnetic state to a conductive and ferromagnetic state upon cooling. In contrast, the LSMO layer in the TT configuration is metallic throughout and exhibits no net magnetization. Such interfacial control of conductivity and magnetization, based on the directions of polarization only, enables a highly versatile architecture for low-voltage magneto-electronic device concepts.

\begin{figure}[htbp!]
  \includegraphics[width=18cm]{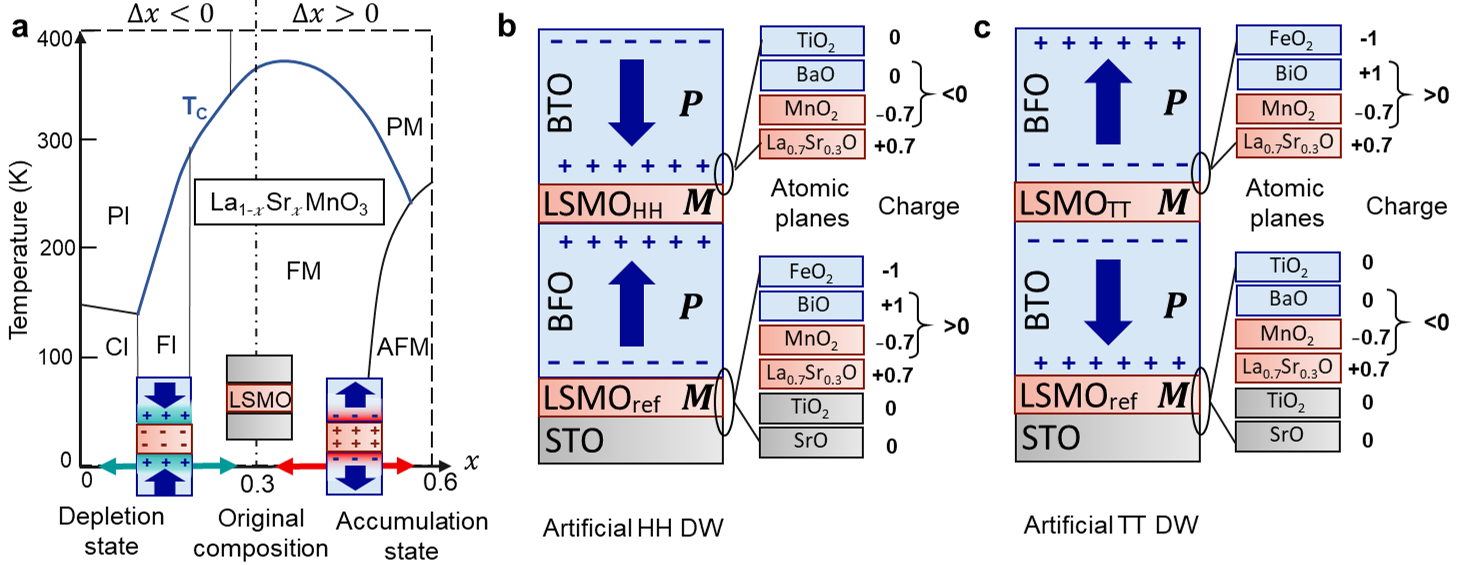}
  \caption{The concept of artificial HH and TT domain walls. (a) A sketch of the La\textsubscript{1-\textit{x}}Sr\textsubscript{\textit{x}}MnO\textsubscript{3} phase diagram as a function of Sr doping ($x$). La\textsubscript{1-\textit{x}}Sr\textsubscript{\textit{x}}MnO\textsubscript{3} layers of original composition $x=0.3$ (LSMO) are sandwiched between two ferroelectric layers with opposite polarization directions. The HH polarization configuration in the trilayer results in electron-doped LSMO\textsubscript{HH} (green shading), comparable to reduced Sr doping ($x<0.3$). The TT polarization configuration in the trilayer leads to the hole-doping in the LSMO\textsubscript{TT} (red shading) comparable to increased Sr doping ($x>0.3$). Abbreviations that denote phases: PM - paramagnetic metal, PI - paramagnetic insulator, FM - ferromagnetic metal, FI - ferromagnetic insulator, AFM - antiferromagnetic metal, CI - spin-canted insulator. Adapted with permission\cite{Dagotto2001} Copyright 2001, Elsevier Science B.V. (b-c) Sketches of heterostructures containing artificial (b) HH and (c) TT domain walls. LSMO\textsubscript{ref} as buffer electrode provides electrostatic screening and additionally serves as reference layers for comparison of magnetic/conduction properties with the LSMO\textsubscript{HH} and LSMO\textsubscript{TT} layers. Atomic planes and their charges, as well as the polarization directions with the corresponding bound charges, are depicted.}
\end{figure}


The tunability between ferromagnetic metallic and antiferromagnetic insulating behavior in the colossal magnetoresistive manganites\cite{Dagotto2001a,Tokura2006} motivates our choice of the La\textsubscript{1-\textit{x}}Sr\textsubscript{\textit{x}}MnO\textsubscript{3} system as the ferromagnetic metal to be inserted into the HH and TT configurations of the ferroelectric layers. Figure 1a shows how different chemical composition of La\textsubscript{1-\textit{x}}Sr\textsubscript{\textit{x}}MnO\textsubscript{3} enables navigation within its phase diagram\cite{Dagotto2001a,Tokura2006}. Variation in Sr content ($x$) gives rise to drastic changes in conduction, accompanied by the emergence of ferromagnetism or antiferromagnetism. Most importantly, electrostatic doping with holes or electrons emulates a change in chemical substitution ($\Delta x>0$ or $\Delta x<0$, respectively). Here we induce such hole/electron doping in La\textsubscript{1-\textit{x}}Sr\textsubscript{\textit{x}}MnO\textsubscript{3} by interfacing it with ferroelectric layers in our functional trilayers. This gives rise to the active and reversible control of the La\textsubscript{1-\textit{x}}Sr\textsubscript{\textit{x}}MnO\textsubscript{3} conductivity and magnetization\cite{Molegraaf2009a,Burton2009a}, set by the orientation of the spontaneous polarization in the adjacent ferroelectric layers. Effectively, the trilayers with the HH polarization configuration correspond to electron doping (depletion state), whereas hole doping (accumulation state) is achieved with the TT configuration\cite{Molegraaf2009a, Vaz2010, Vaz2011} (Figure 1a). Such electrostatic doping at the interfaces with ferroelectric layers affects only a few La\textsubscript{1-\textit{x}}Sr\textsubscript{\textit{x}}MnO\textsubscript{3} unit cells\cite{Hong2005a,Rondinelli2008, Burton2009a}. However, if the La\textsubscript{1-\textit{x}}Sr\textsubscript{\textit{x}}MnO\textsubscript{3} films are ultrathin, this local interfacial effect can dominate the overall response of the heterostructures\cite{Spurgeon2014a}. We select La\textsubscript{1-\textit{x}}Sr\textsubscript{\textit{x}}MnO\textsubscript{3} at $x=0.3$ (LSMO), as this composition grants a high Curie temperature (\textit{T}\textsubscript{C}) and potentially enables phase transitions upon both electron doping (e.g. FM$ \rightarrow$PI) and hole doping (e.g. FM$ \rightarrow$PM). Note that in contrast to direct Sr substitution in bulk LaMnO\textsubscript{3} crystals, the interfacial electrostatic doping in thin films cannot be quantified in terms of the corresponding shift in $x$ because factors such as orbital ordering imposed by epitaxial strain\cite{Fang2000} and dissimilar polar discontinuities at interfaces\cite{Boschker2012b} exert additional influence on the LSMO conductivity and magnetization. 

In order to achieve ferroelectric layers with deterministic polarization directions, we make use of the mismatch in layer charges at the interfaces of our heterostructures\cite{rijnders04,Yu2012a, Guo2016,DeLuca2017a}. \textit{AB}O\textsubscript{3}-type perovskite materials contain \textit{A}O and \textit{B}O\textsubscript{2} atomic planes that can be either neutral or charged depending on the choice of the \textit{A} and \textit{B} cations. If the ferroelectric and the buffer layer beneath possess atomic planes with dissimilar charge, this gives rise to a charge discontinuity at the interface. A surplus of positive charges can screen negative bound charges, triggering an upward-oriented polarization in the ferroelectric, whereas an excess of negative charges forces the polarization to point downward. To achieve opposing polarization directions on a fixed, charged buffer layer, two ferroelectrics with differently charged atomic planes must be selected. For an MnO\textsubscript{2}\textsuperscript{$-$0.7}-terminated LSMO buffer on an STO substrate, the pairing of BFO and BTO matches this criterion. The interface between LSMO and BFO has a net positive charge, favoring upward polarization in BFO, whereas the interface between LSMO and BTO has a net negative charge, inducing downward polarization in BTO; see Figure 1b,c. In this way, we obtain HH and TT polarization configurations simply by inverting the BFO-LSMO-BTO growth sequences in our heterostructures, as shown in Figure 1b,c.  

Note that the 15-unit-cell (u.c.)-thick LSMO enters our heterostructures in two ways: (i) as domain-wall element (LSMO\textsubscript{HH} and LSMO\textsubscript{TT}), (ii) as buffer electrode (LSMO\textsubscript{ref}), which provides electrostatic screening, sets the MnO\textsubscript{2}\textsuperscript{$-$0.7} termination, and additionally serves as a reference layer for comparison of magnetic/conduction properties. The LSMO thickness of 15 u.c. ensures that the interfacial charge-coupling effects dominate the net magnetic response of the heterostructures. 




\section{RESULTS AND DISCUSSION}


We grew our BTO\(|\)LSMO\(|\)BFO (HH) and BFO\(|\)LSMO\(|\)BTO (TT) heterostructures by pulsed laser deposition on LSMO-buffered, TiO\textsubscript{2}-terminated SrTiO\textsubscript{3} (STO) (001) substrates. In order to follow the polar state of each layer during the epitaxial design, we use ISHG. Optical SHG is a nonlinear optical process sensitive to inversion symmetry breaking by long-range order. When used to monitor ferroelectric thin films, the amplitude of the SHG wave is indicative of the net polarization in the probed material\cite{DeLuca2017a, Nordlander2018}. It even provides us with information about polar states that are buried and hence inaccessible with techniques like PFM. The ISHG signal is measured in a 45\textdegree{} reflection geometry. The tilted incidence permits us to detect the relevant out-of-plane component of the polarization in our heterostructures associated with the artificial HH and TT domain walls\cite{DeLuca2017a}. In addition, reflection high-energy electron diffraction (RHEED) is used to simultaneously monitor the thickness of the layers, to ensure the desired surface termination, and to record our ISHG yield with unit-cell accuracy. 

The ISHG signal collected during the growth of heterostructures corresponding to the HH and TT polarization configurations is shown in Figure 2a,b, respectively. In both cases, the ISHG signal evolves as follows. During the growth of the first (bottom) ferroelectric layer the ISHG intensity is continuously increasing. Subsequent deposition of the functional LSMO layer results in the signal dropping to the background level. During the deposition of the second (top) ferroelectric layer, the ISHG intensity re-emerges, continuously increases, and remains stable when the growth is halted.

Such in-situ access to the polarization throughout the entire multilayer deposition process gives crucial information on the polarization configuration in our heterostructures. The slight delay in the onset of the ISHG signal during the growth of all ferroelectric layers is a clear indication of a critical thickness for the emergence of the spontaneous polarization, caused by the limited screening efficiency of the metallic LSMO buffer\cite{Chu2007a, DeLuca2017a}. Note that the difference in ISHG yield obtained for BFO and BTO layers is related to the different magnitude and tensor components of the SHG susceptibility parametrizing the nonlinear light-matter interaction\cite{Nordlander2018} as well as to dissimilar values of the spontaneous polarization associated with the two compounds. 

The drop in ISHG intensity during the LSMO deposition observed for the HH- and TT-type heterostructures cannot be explained by light absorption only but is attributed to the formation of out-of-plane-polarized domains in the ferroelectric bottom layer. This is caused by the low charge-screening efficiency of LSMO at its early growth stage\cite{Huijben2008}. The resulting depolarizing field initiates multi-domain formation\cite{Strkalj2019}. As the net polarization decreases, so does the ISHG intensity. The re-emergence of the ISHG signal with the completion of the second ferroelectric layer is consistent with a stabilization of a single-domain configuration in this layer, as we confirmed post-deposition by PFM. 


\begin{figure} \centering
  \includegraphics[width=16.6cm]{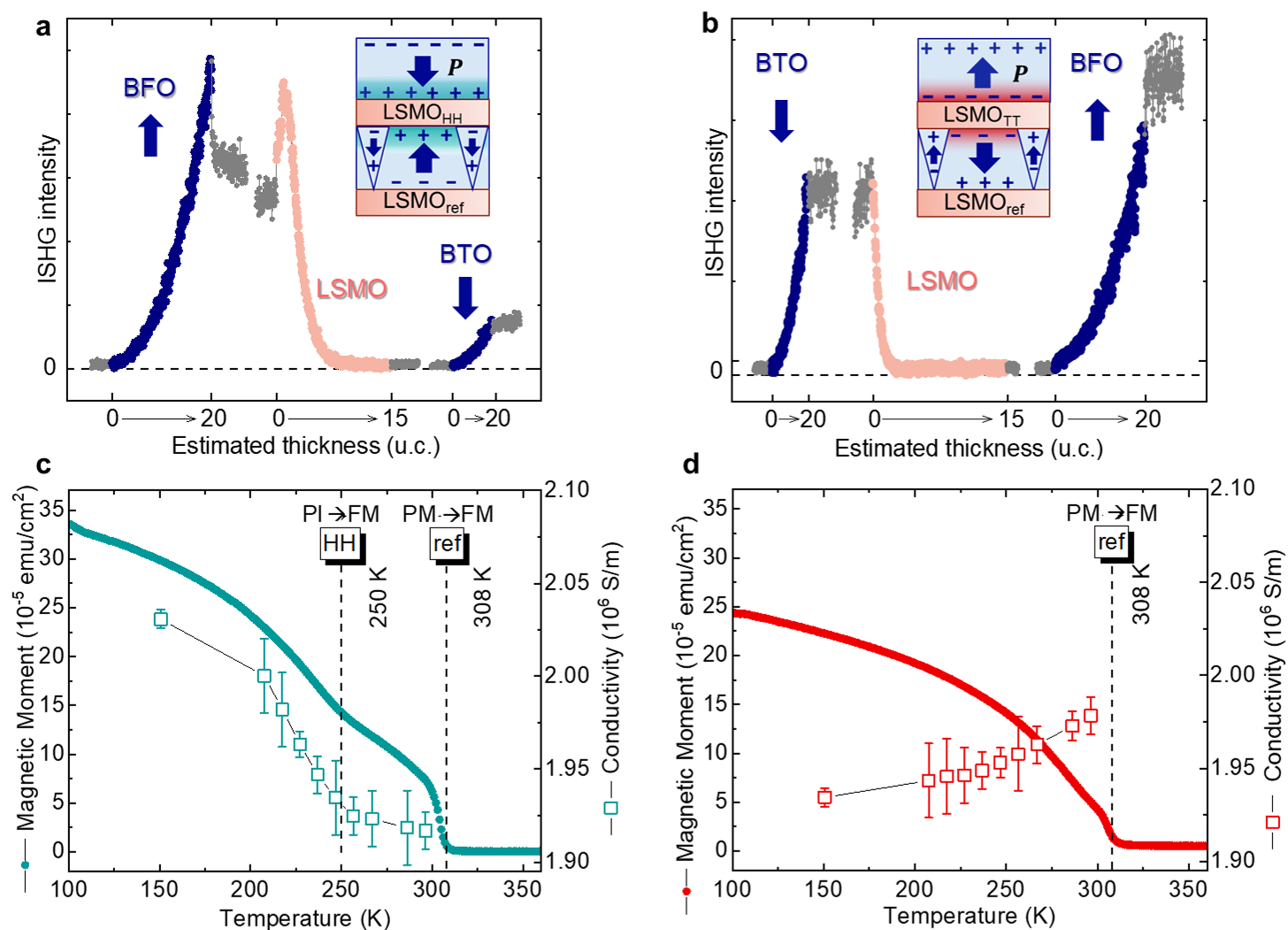}
  \caption{Design of heterostructures with artificial HH and TT domain walls and comparison of their properties. (a-b) ISHG real-time monitoring during the growth of the heterostructures with artificial (a) HH and (b) TT domain walls. Gray data points show the ISHG intensity recorded for two minutes prior to and after the deposition of each layer. The insets show the polarization configurations in both heterostructures and the resulting interfacial doping of the LSMO\textsubscript{HH} and LSMO\textsubscript{TT}: electron doping (green shading) and hole doping (red shading). (c-d) Temperature dependence of magnetization and optical conductivity (error bars correspond to the standard deviation) measured for the heterostructures with artificial (c) HH and (d) TT domain walls.Temperatures corresponding to the observed phase transitions PM$\,\to\,$FM and PI$\,\to\,$FM (see Figure 1a) are marked. We note that the sharp decline and subsequent rise in the ISHG signal in the time window between the BFO and LSMO deposition in (a) is related to interface proximity effects in ultrathin ferroelectrics\cite{Strkalj2020}.  
  }
\end{figure}

At first glance, the domain formation occurring in the ferroelectric bottom layer appears to challenge the feasibility of the trilayers emulating charged domain walls between inherently uniformly polarized ferroelectric domains. However, there is some evidence that, despite the observed multi-domain configurations in both TT and HH heterostructures, one polarization state (T or H, respectively) still dominates in the bottom ferroelectric layer, allowing the concept of artificial charged domain walls to be retained. Notably, one clear indication of residual electrostatic doping is seen in the ISHG signal: during BTO deposition, the ISHG signal on LSMO\textsubscript{HH} is significantly weaker than on LSMO\textsubscript{ref} in the TT configuration (see Figure 2a and 2b), despite identical strain states. This suggests a reduced capacity of LSMO\textsubscript{HH} to screen the polarization of the top ferroelectric, likely due to its diminished metallicity. This is compatible with the expected electron doping ($\Delta x < 0$). These findings indicate that LSMO layers are electrostatically influenced by the underlying ferroelectric films, prompting further investigation into the conductivity and magnetization changes in the LSMO junctions in different polarization configurations.

We used SQUID magnetometry to evaluate the magnetization of the LSMO layers and employed non-invasive THz-TDS to extract its optical conductivity directly from the as-grown BTO\(|\)LSMO\(|\)BFO and BFO\(|\)LSMO\(|\)BTO multilayers. Let us first compare the magnetic behavior and conductivity of LSMO\textsubscript{HH} and LSMO\textsubscript{TT}. 

Figure 2c,d show the associated magnetization and optical conductivity measurements as a function of temperature. While the magnetic signature of the LSMO\textsubscript{ref} buffer is present in both samples with an onset of the spontaneous in-plane magnetization at T\textsubscript{C} $\simeq$ 308 K, an additional ferromagnetic transition is visible at 250 K exclusively in the heterostructure with the artificial HH domain wall (Figure 2c). This 250-K transition is accompanied by an increase in conductivity, revealed by the THz-TDS measurements. The combined emergence of conductivity and magnetization during the cool-down suggests the phase transition of LSMO\textsubscript{HH} from PI to FM, consistent with the expected electron doping ($\Delta x < 0$) in relation to the original $x = 0.3$ composition. In contrast, the temperature-dependent magnetic and optical conductivity responses of the LSMO\textsubscript{TT} with the anticipated hole doping ($\Delta x > 0$) show no significant magnetic or conductivity response beyond that of LSMO\textsubscript{ref} down to low temperature (Figure 2d). This is consistent with a phase transition in the LSMO\textsubscript{TT} from PM to AFM in the lower temperature range, which is expected for hole doping $\Delta x > 0$. During this phase transition, neither the magnetic moment of the LSMO\textsubscript{TT} nor its conductivity change, in agreement with our data.

\begin{figure} \centering
  \includegraphics[width=12cm]{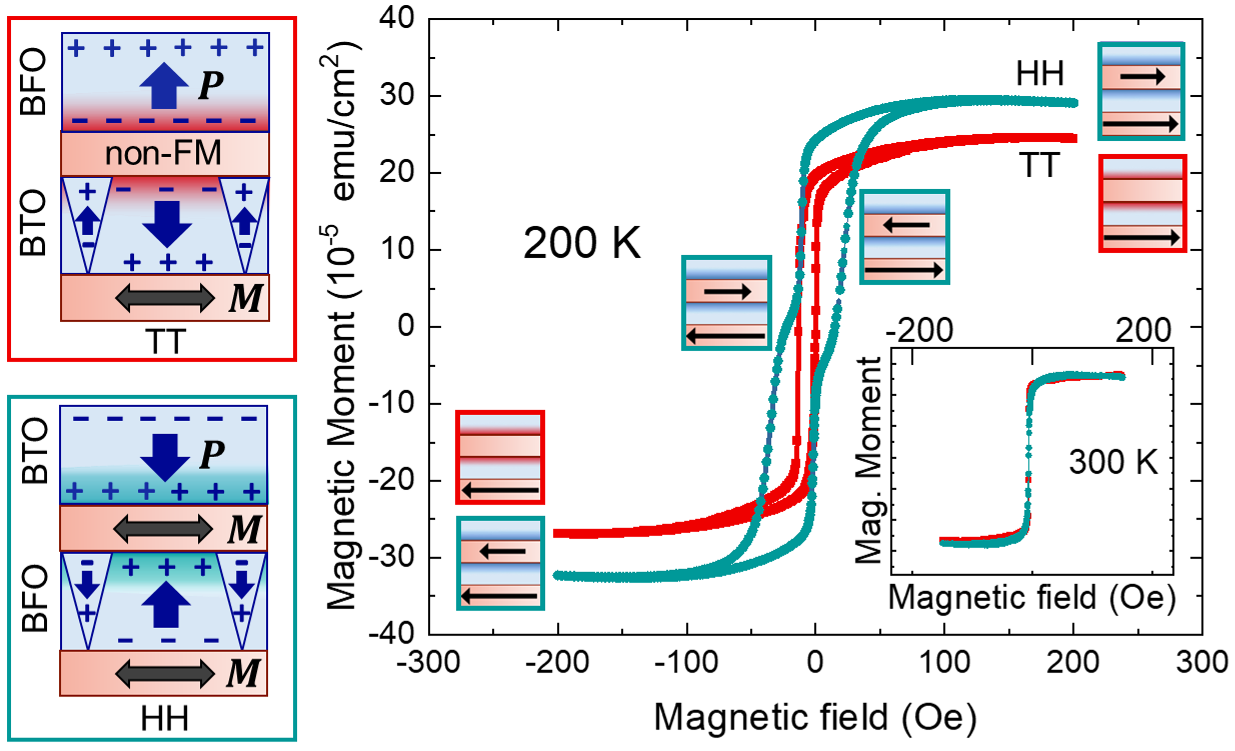}
  \caption{Magnetic hysteresis curves of heterostructures with artificial TT (red) and HH (green) domain walls recorded at 200 K and 300 K (inset). Schematics show the heterostructures and the associated magnetization state of the LSMO layers (black arrows) at 200 K. H\textsubscript{1} and H\textsubscript{2} indicate two coercive fields observed for the heterostructure with the HH artificial domain wall at 200 K.}
\end{figure}

To highlight the functionality of the created artificial domain walls, in Figure 3 we show magnetic-field-dependent magnetization curves for the HH- and TT-like heterostructures. The room-temperature magnetization curves (see inset of Figure 3) allow us to identify and isolate the common magnetic response originating from the bottom LSMO\textsubscript{ref} electrode since the LSMO\textsubscript{HH} and LSMO\textsubscript{TT} layers are paramagnetic at this temperature. At 200 K, we obtain a double hysteresis with two superimposed magnetic data for the heterostructure with the artificial HH domain wall: one of higher coercivity (32 Oe, H\textsubscript{2}) and lower magnetization compared to the other (7 Oe, H\textsubscript{1}). The softer magnetic contribution associated with H\textsubscript{1} in the HH-like heterostructure is very similar to the one of the TT case and can therefore be attributed to the reference bottom electrode LSMO\textsubscript{ref}. In contrast, the harder magnetic response associated with H\textsubscript{2} and the lower magnetic moment originate from the LSMO\textsubscript{HH}. The difference in magnetic coercivity between the LSMO\textsubscript{ref} and LSMO\textsubscript{HH} can be rationalized by the gradient in charge doping across the LSMO\textsubscript{HH} thickness. The electrostatic doping primarily affects the LSMO\textsubscript{HH} surfaces and may not affect the entire thickness of the film. Here, at 200 K, we argue that the surfaces of LSMO\textsubscript{HH} exhibit a higher degree of electron doping and remain in the PI phase. The inner volume of the LSMO\textsubscript{HH}, in contrast, adopts the FM order. This results in a reduced FM thickness in the LSMO\textsubscript{HH} compared to LSMO\textsubscript{ref}, which is commonly associated with higher magnetic coercivity in ultrathin LSMO films\cite{Huijben2008} (Here, 32 Oe versus 7 Oe). The LSMO\textsubscript{TT} layer exhibits no magnetic signature at this temperature, corroborating the inferred antiferromagnetic ordering.

The results presented thus far demonstrate the control of conductivity and magnetization in our model-kit HH and TT domain walls, confirming interfacial magnetoelectric coupling. However, a significant challenge we encountered was the multi-domain breakdown in the bottom (capped) ferroelectric layer, which decreased the charge density available for electrostatic doping of the lower LSMO interface. By finding ways to avoid this issue and stabilizing a single-domain configuration in both ferroelectric layers, we might be able to enhance charge accumulation at the functional LSMO layers even further, thereby boosting the magnetoelectric coupling in our heterostructures.

A uniform out-of-plane polarization in capped ferroelectric thin films can be stabilized by simultaneous tailoring of the electrostatics at the top and bottom interfaces\cite{Strkalj2020, Gattinoni2020}. In the case of BFO, the notorious bismuth volatility triggers the formation of a positively charged off-stoichiometric surface layer\cite{Alexe1998,Ba2005} favoring a downward polarization. To obtain a downward rather than upward-oriented polarization in BFO, we change the charged-plane termination in the LSMO\textsubscript{ref} buffer. This approach ensures that both top and bottom surfaces contribute cooperatively, stabilizing a single-domain configuration in the BFO film\cite{Strkalj2020, Gattinoni2020}. To achieve this, we enforce (La,Sr)O termination of the LSMO buffer by inserting an SrO-terminated SrRuO\textsubscript{3} (SRO) layer of 2 u.c.\ between the substrate and the bottom LSMO\textsubscript{ref} buffer, see Figure 4a. This changes the surface termination of the layers in the heterostructure from \textit{A}O to \textit{B}O\textsubscript{2}, which leads to the consequent inversion of the polarization direction in the ferroelectric layers.


\begin{figure} \centering
  \includegraphics[width=14cm]{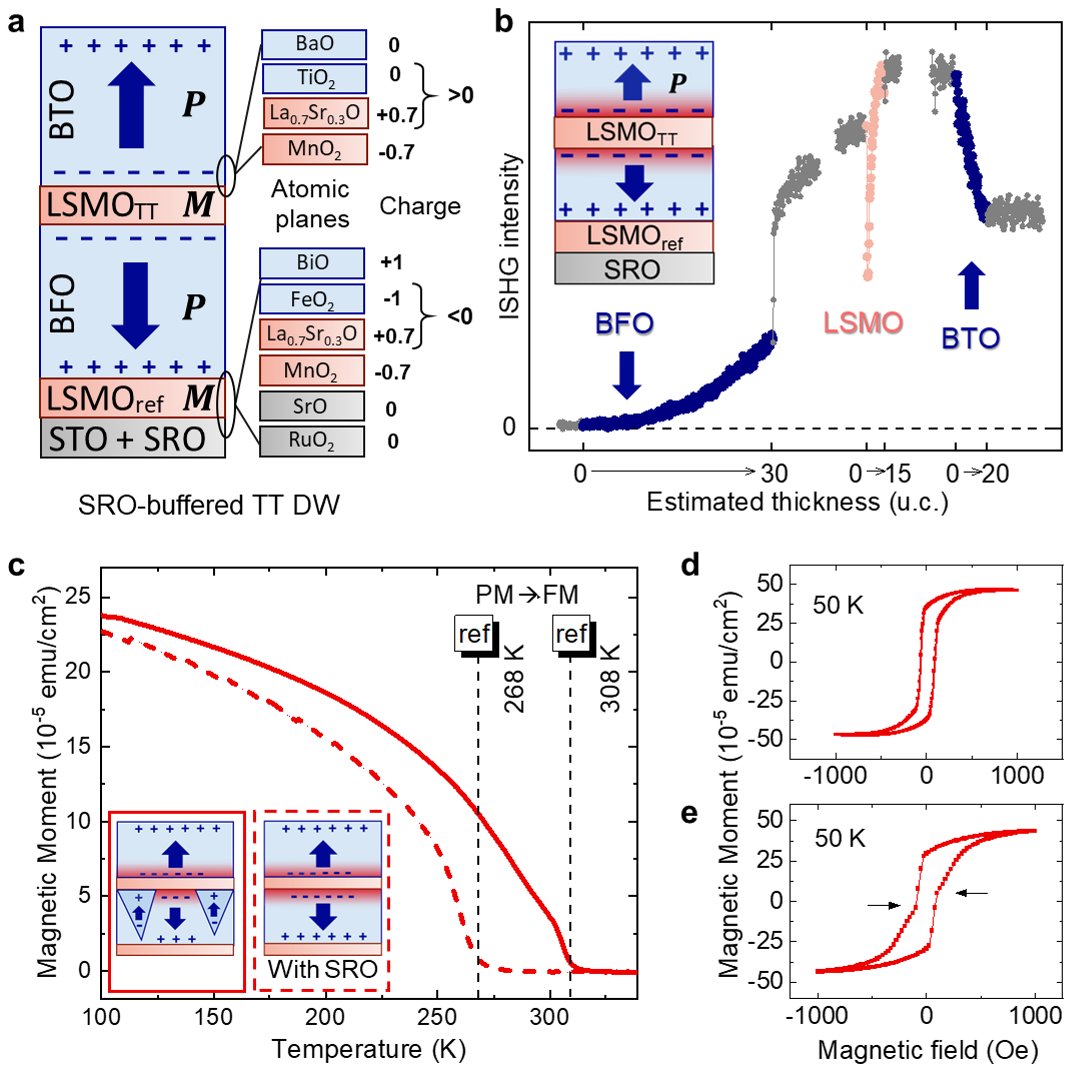}
  \caption{Enhancing charge coupling by inhibiting domain formation in the lower ferroelectric layer. 
(a) The heterostructure with the artificial TT domain wall achieved with BTO\(|\)LSMO\(|\)BFO and the insertion of a 2-u.c.\-thick SRO interlayer. (b) ISHG monitoring during the growth of the heterostructure with the artificial TT domain wall buffered with the SRO interlayer. Gray data points show the ISHG intensity recorded for two minutes prior to and after the deposition of each layer. (c) Temperature-dependent magnetization in heterostructures with single- and multi-domain configurations in the buried ferroelectric layer. (d-e) Magnetic hysteresis curves at 50 K for heterostructures with an artificial TT domain wall: (d) with a single-domain configuration and (e) with a multi-domain configuration in the lower ferroelectric layer. Arrows in (e) highlight changes in the magnetic response characteristic of two different ferromagnetic contributions.}
\end{figure}


Figure 4b displays the ISHG data collected during the growth of the downward-oriented BFO.  This time, we observe a more than two-fold enhancement of the ISHG signal that ensues after the completion of BFO deposition (see grey data points after the BFO deposition). This enhancement of polarization is consistent with the coordinated interfacial electrostatic contributions\cite{Strkalj2020} that promote a downward-oriented polarization in the BFO. During subsequent capping with the LSMO layer, the ISHG signal briefly dips during the deposition of the first few LSMO unit cells due to their low charge-screening potential\cite{Huijben2008} that momentarily affects the BFO polarization. However, once the metallicity in LSMO is established, we see a recovery in the ISHG signal, compatible with the single-domain configuration in the capped BFO layer. In contrast to our previous design of heterostructures that led to multi-domain bottom ferroelectric layers (see Figure 2a,b), here, the consecutive deposition of BTO now gives rise to a decreasing ISHG signal. This is a signature of destructive interference\cite{Sarott2021} between the now non-zero ISHG signal originating from the buried, uniformly polarized BFO layer and the ISHG signal emitted from the upward-polarized BTO. We thus conclude that we have successfully addressed the remaining shortcoming of our domain-wall-like architecture and achieved the artificial TT domain wall, with all ferroelectric layers in a single-domain configuration and exhibiting opposing out-of-plane polarization.

In Figure 4c we first compare the magnetic response of the two heterostructures with artificial TT domain walls, grown without and with the SRO interlayer. Both heterostructures show only one phase transition associated with the LSMO\textsubscript{ref} layer (from PM to FM), which indicates that the antiferromagnetic order of the LSMO\textsubscript{TT} layer is maintained in the new SRO-buffered heterostructure. This demonstrates that the magnetic properties of LSMO\textsubscript{TT} are preserved despite the reversed atomic surface termination and the exchanged deposition sequence of the ferroelectric compounds in the multilayer. 

Let us now examine how the stabilized single-domain configuration in the bottom ferroelectric layer of our SRO-buffered TT multilayer influences the magnetic response of LSMO\textsubscript{TT}. Comparing the hysteresis loops in Figures 4d and 4e reveals that the preserved single-domain configuration in the lower ferroelectric layer enforces an AFM ordering in LSMO\textsubscript{TT} over a larger temperature range compared to the heterostructure with the multi-domain bottom ferroelectric layer. At 50 K, only one contribution to the hysteresis loop, related to LSMO\textsubscript{ref}, is observed (Figure 4d). In contrast, the heterostructure with a bottom multi-domain ferroelectric layer exhibits two distinct ferromagnetic signals at this temperature, highlighted by arrows in Figure 4e. This suggests that the higher degree of hole doping in the heterostructure with a single-domain ferroelectric layer ensures that antiferromagnetic ordering in LSMO\textsubscript{TT} is maintained down to a lower temperature. Conversely, in the heterostructure with the multi-domain ferroelectric layer, and consequently more moderate hole doping, LSMO\textsubscript{TT} shows signs of gradually transitioning into ferromagnetic ordering at low temperatures. This confirms that uniform polarization in the bottom BFO layer increases hole doping. Additionally, the single-domain bottom ferroelectric layer also leads to higher electron doping of the LSMO\textsubscript{ref} buffer layer, reflected in the lowering of its T\textsubscript{C} value from 307 K to 268 K, as seen in Figure 4c. Thus, we demonstrate that the properties of our artificial magnetoelectric domain walls are predominantly driven by electrostatic doping at the interfaces with ferroelectric layers rather than other effects (like interfacial chemistry), highlighting the generality of our proposed concept of artificial magnetoelectric domain walls.

\section{CONCLUSIONS}
With this work, we introduce the concept of artificial domain walls with HH and TT charge configurations. By utilizing out-of-plane-polarized BFO and BTO ferroelectric thin films, we engineer interfaces that are either positively or negatively charged. To translate these electron and hole accumulations into tunable functionalities, we insert ultrathin LSMO junctions that are sensitive to charge doping. We demonstrate that the conductivity and magnetization of these artificial domain walls can be selectively controlled through interfacial magnetoelectric coupling, depending on whether the HH or TT polarization configuration is adopted in the multilayer structure. At room temperature, the electron-doped LSMO layer, which emulates an HH domain wall, is insulating and paramagnetic, but transitions to conductive and ferromagnetic as the temperature decreases. In contrast, the hole-doped LSMO layer, mimicking a TT domain wall, is metallic and paramagnetic at room temperature and transitions to an antiferromagnetic metallic phase at lower temperatures.

ISHG experiments provide in-situ access to polarization throughout the entire multilayer deposition process, offering crucial information on the polarization configuration in our heterostructures. This enables us to construct heterostructures with ferroelectric layers in single-domain configurations, maximizing charge coupling at the interfaces of our artificial domain walls and improving their functionality over a wider temperature range. This research advances our understanding of electrical control over magnetism and conductivity in ultrathin multiferroic oxide heterostructures, in the ongoing quest for tunable, energy-efficient oxide electronics.


\section*{Experimental Section}
\RaggedRight

\subsection{Heterostructure Growth}
Heterostructures were grown on TiO\textsubscript{2}-terminated STO (001) substrates (CrysTec GmbH) by pulsed laser deposition using a 248 nm KrF excimer laser. The deposition was performed at a substrate temperature of 700$^{\circ}$C under an oxygen partial pressure of 0.10 mbar (SRO) or 0.15 mbar (other layers) with a laser fluence of 0.9 J cm\textsuperscript{−2}. The thickness of the thin films was monitored using a combination of RHEED during growth and X-ray reflectivity ex-situ. LSMO layers were 15 u.c. thick allowing for interfacial charge-coupling effects to dominate the bulk response of the heterostructures. The coherent strain in each heterostructure was confirmed by reciprocal space mapping measurements around the STO (103) reflection, using a four-circle X-ray diffractometer (Panalytical X'Pert\textsuperscript{3} MRD). To confirm the polarization direction of each ferroelectric constituent layer, PFM acquisition was performed on the reversibly poled areas using a 2-V peak-to-peak AC modulation at 69 kHz with a scanning probe microscope (NT-MDT NTEGRA, Spectrum Instruments). \\

\subsection{ISHG monitoring} The optical SHG signal was generated in-situ in 45\textdegree{} angle of incidence to the sample in the PLD growth chamber\cite{DeLuca2017a}. The output of an amplified Ti:Sapphire laser system (wavelength: 800 nm, repetition rate: 1 kHz, pulse duration: 45 fs) was converted by an optical parametric amplifier into the fundamental light with a wavelength of 1200 nm.  This probe beam was incident on the sample with a pulse energy of 10 \SI{}{\micro\joule} on a spot size 250 \SI{}{\micro\metre} in diameter. The generated ISHG intensity was detected using a monochromator (Triax, Horiba) set to 600 nm and a photomultiplier system. Both the incident light and the detected ISHG light are polarized parallel to the plane of reflection.  \\
\subsection{Magnetic Characterization} All the magnetic measurements were conducted using VSM-SQUID (MPMS3, Quantum Design) with a magnetic field applied in the film plane. For the temperature-dependent magnetization measurements, samples were demagnetized at \SI{380}{\kelvin}, and then the change of magnetic moment was measured upon cooling in a field of 20 Oe applied in the film plane. \\
\subsection{Optical Conductivity Measurements} THz conductivity is measured using the time-domain THz spectroscopy in the reflection geometry. Single-cycle terahertz pulses are generated by optical rectification in a 0.5-mm ZnTe(110)-oriented single crystal, using 90\% of a Ti:Sapphire laser output (wavelength 800 nm, pulse duration 120 fs, pulse repetition rate 1 kHz, pulse energy 2 mJ). The remaining 10\% of the fundamental beam is used as a gating pulse for the free-space electrooptic sampling of the reflected THz wave. The THz and the gating beams are collinearly focused onto a ZnTe(110)-oriented detection crystal. The THz-induced ellipticity of the gating pulse is measured using a quarter-wave plate, a Wollaston polarizer, and a balanced photodiode. The signal from the photodiode is then analyzed with a lock-in amplifier. All temperature-dependent measurements are performed in an inert nitrogen atmosphere. \\

\newpage

\section*{References}

\bibliographystyle{ieeetr}
\bibliography{ADW2}

\begin{addendum}
\item[Author contributions] All authors discussed the results. E.G. and M.T. wrote the manuscript with M.F. The thin-film growth, the ISHG measurements as well as the structural, electric, and magnetic characterization were performed by E.G. Optical conductivity measurements were carried out by C.-J.Y. and S.P. The experiments were designed by M.T. and supervised by M.T. jointly with M.F.

\item[Acknowledgements] E.G. and M.T. acknowledge the Swiss National Science Foundation under Project No. 200021\_188414. M.T. and M.F. acknowledge support by the EU European Research Council under Advanced Grant Program No. 694955-INSEETO. E.G. acknowledges the Swiss National Science Foundation under Project No. P500PT\_214449. S.P. acknowledges the support from ETH Career Seed Grant No. SEED-17 18-1 and SERB through SERB-SRG via Project No. SRG/2022/000290.

\item[Competing Interests] The authors declare that they have no competing financial interests.

\end{addendum}

\end{document}